\documentclass{article}

\usepackage{natbib}
\usepackage{graphicx,xspace}
\usepackage[acronym,xindy,toc,nonumberlist,shortcuts]{glossaries} 
\usepackage{url}

\usepackage[linesnumbered,ruled,vlined]{algorithm2e}
\usepackage{floatrow}
\usepackage{enumitem}
\usepackage{graphics}
\usepackage{amsmath}
\usepackage{sectsty}
\sectionfont{\bfseries}
\subsectionfont{\normalsize}

\usepackage{array}
\newcolumntype{P}[1]{>{\raggedright\arraybackslash}p{#1}}

\usepackage{todonotes}

\usepackage{color,soul} % highlight text
\definecolor{lightgreen}{rgb}{.8,1,.8}
\definecolor{darkgreen}{rgb}{.6,1,.6}
\sethlcolor{lightgreen}
\setstcolor{red}

\usepackage[normalem]{ulem}
\DeclareRobustCommand\reduline{\bgroup\markoverwith{\textcolor{red}{\rule[0.5ex]{2pt}{0.4pt}}}\ULon}
\DeclareRobustCommand\greenuline{\bgroup\markoverwith{\textcolor{darkgreen}{\rule[-0.9ex]{4pt}{3pt}}}\ULon}
\DeclareRobustCommand{\changed}[1]{#1} % soul
\DeclareRobustCommand{\removed}[1]{} % Hide all removed text

\newacronym{lod}{LOD}{linked open data}
\newacronym{vgi}{VGI}{volunteered geographic information}
\newacronym{osm}{OSM}{OpenStreetMap}
\newacronym{hci}{HCI}{human-computer interaction}
\newacronym{ai}{AI}{artificial intelligence}
\newacronym{gis}{GIS}{geographic information system}
\newacronym{idf}{IDF}{Inverse Document Frequency}
\newacronym{lsa}{LSA}{Latent Semantic Analysis}
\newacronym{poipl}{POI}{points of interest}
\newacronym{poi}{POI}{point of interest}
\newacronym{vsm}{VSM}{vector space model}
\newacronym{pos}{POS}{part-of-speech}
\newacronym{gwap}{GWAP}{game with a purpose}
\newacronym{www}{WWW}{World Wide Web}
\newacronym{skos}{SKOS}{W3C Simple Knowledge Organization System}
\newacronym{rdf}{RDF}{Resource Description Framework}
\newacronym{owl}{OWL}{Web Ontology Language}
\newacronym{oaei}{OAEI}{Ontology Alignment Evaluation Initiative}
\newacronym{giscience}{GIScience}{geographic information science}
\newacronym{wsddef}{WSD}{word sense disambiguation}
\newacronym{dl}{DL}{description logic}
\newacronym{mdsmsim}{MDSM}{Matching-Distance Similarity Measure}
\newacronym{oursurvey}{GeReSiD}{Geo Relatedness and Similarity Dataset}
\newacronym{ira}{IRA}{interrater agreement}

\newacronym{irr}{IRR}{interrater reliability}
\newacronym{mds}{MDS}{multidimensional scaling}
\newacronym{lbs}{LBS}{location-based services}
\newacronym{oss}{FOSS}{free and open-source software}
\newacronym{sdts}{SDTS}{Spatial Data Transfer Standard}
\newacronym{tfidf}{TF-IDF}{Term Frequency-Inverse Document Frequency}
\newacronym{ogc}{OGC}{Open Geospatial Consortium}

\newglossaryentry{w2}{name=Web 2.0,
	description={TODO}
}

\newglossaryentry{bow}{name=bag-of-words}
\newglossaryentry{gir}{name=geographic information retrieval}
\newglossaryentry{ir}{name=information retrieval}
\newglossaryentry{gn}{name=GeoNames}
\newglossaryentry{gwn}{name=GeoWordNet}
\newglossaryentry{gno}{name=GeoNames ontology}
\newglossaryentry{algo}{name=\emph{Voc2WordNet}}
\newglossaryentry{algoplain}{name=Voc2WordNet}

\newglossaryentry{lodcloud}{name=LOD cloud}

\newglossaryentry{stratag}{name=Strategic Research in Advanced Geotechnologies (StratAG)}

\newglossaryentry{wn}{name=WordNet}

\newglossaryentry{kb}{name={knowledge base}}
\newglossaryentry{gkb}{name={geo-knowledge base}}

\newglossaryentry{nuim}{name={National University of Ireland, Maynooth},
	description={TODO}
}

\newglossaryentry{ucd}{name={University College Dublin},
	description={TODO}
}

\newglossaryentry{osmsim}{name=\textsc{OSM-TagSim},
	description={TODO}
}

\newglossaryentry{lexsim}{name=\textsc{osm-sim_{lex}},
	description={TODO}
}

\newglossaryentry{simdl}{name=Sim-DL,
	description={TODO}
}

\newglossaryentry{netsim}{name=\textsc{osm-sim_{sim}},
	description={TODO}
}

\newglossaryentry{dbp}{name=DBpedia,
	description={TODO}
}

\newglossaryentry{lgd}{name=LinkedGeoData,
	description={TODO}
}

\newglossaryentry{sgw}{name=Semantic Geospatial Web,
	description={TODO}
}

\newglossaryentry{sw}{name=Semantic Web,
	description={TODO}
}

\newglossaryentry{webplatform}{name=Web platform for map personalisation and visualisation,
	description={TODO}
}

\newglossaryentry{wsd}{name=word sense disambiguation,
	description={TODO}
}

\newglossaryentry{os}{name=open source,
	description={TODO}
}

\newglossaryentry{osn}{name=OSM Semantic Network,
	description={TODO}
}

\newglossaryentry{nlp}{name=natural language processing,
	description={TODO}
}

\newglossaryentry{owc}{name=OSM Wiki Crawler,
	description={TODO}
}

\newglossaryentry{oww}{name=OSM Wiki website,
	description={\url{http://wiki.openstreetmap.org}}
}

\newglossaryentry{mdsm}{name=MDSM evaluation dataset,
	description={TODO}
}

\newcommand{\accesseddate}[0]{(acc. Oct 30, 2012)}

\newcommand{\osmtag}[1]{\emph{#1}} 
\newcommand{\footurln}[2]{\footnotemark\footnotetext{#1 \url{#2}}}
\newcommand{\footurl}[1]{\footnotemark\footnotetext{\url{#1}}}
\newcommand{\footurltwo}[2]{\footnotemark\footnotetext{\url{#1}, \url{#2}}}
\newcommand{\footurlthree}[3]{\footnotemark\footnotetext{\url{#1}, \url{#2}, \url{#3} \accesseddate{}}}

\newcommand{\homepage}[1]{\url{http://github.com/ucd-spatial/#1}}
\newcommand{\semuri}[1]{\emph{#1}}
\newcommand{\shortu}[1]{{\url{#1}}}

 \date{\small{Author copy. Published in \emph{Annals of GIS}, 20 (2) 2014}}
\title{Linking Geographic Vocabularies\\through WordNet}
 
\author{A. Ballatore,\thanks{School of Computer Science and Informatics, University College Dublin, Ireland. \texttt{andrea.ballatore@ucd.ie}} $~$ M. Bertolotto,\thanks{School of Computer Science and Informatics, University College Dublin, Ireland.}$~$ and D.C. Wilson\thanks{Department of Software and Information Systems, University of North Carolina, Charlotte, NC}}

\begin{document}

\maketitle

\begin{abstract}
\noindent
The linked open data paradigm has emerged as a promising approach to structuring and sharing geospatial information.
One of the major obstacles to this vision lies in the difficulties found in the automatic integration between heterogeneous vocabularies and ontologies that provides the semantic backbone of the growing constellation of open \glspl{gkb}.
%\Gls{osm}, for example, has gathered billions of geographic objects from its contributors in a spatial folksonomy.
In this article, we show how to utilise \gls{wn} as a semantic hub to increase the integration of linked open data.
With this purpose in mind, we devise \gls{algo}, an unsupervised mapping technique between a given vocabulary and \gls{wn}, combining intensional and extensional aspects of the geographic terms.
\gls{algo} is evaluated against a sample of human-generated alignments with the OpenStreetMap Semantic Network, a crowdsourced geospatial resource, and the \gls{gno}, the vocabulary of a large digital gazetteer.
These empirical results indicate that the approach can obtain high precision and recall.

\end{abstract}

%=============================================
\section{Introduction}
\label{sec:intro}
%=============================================
\glsresetall

% why LOD
%Since its invention in 1990, the World Wide Web has enabled an unprecedented growth of digital data, offering a platform for publishing, retrieving, and sharing any type of data across the globe.
Over the past decades, a large volume of \removed{geospatial} \changed{digital} information has been disseminated online in a variety of incompatible formats and heterogeneous data spaces.
This semantic gap hinders the ability to analyse, explore, and discover unexpected connections and relations between entities, obtaining insights about complex social, geographic, cultural, and economic processes.
% key tenets of LOD
Berners-Lee's \gls{sw} is a prominent attempt to overcome this crucial gap, and to provide a flexible and yet unified platform for data sharing \citep{Berners:2001:semantic}.
One of the most promising initiatives in this ambitious framework is the so-called \emph{\gls{lod}} paradigm, with the purpose of creating a unified data space.
To be classified as \gls{lod}, data must be (i) released under open licenses; (ii) saved in a machine-readable digital format; (iii) stored in non-proprietary formats; (iv) accessible via URIs; and (v) linked to other \gls{lod}.\footnote{\url{http://5stardata.info} - All URLs cited were accessed on \today.}
As \gls{lod} is generated and published online, a graph of datasets has emerged, resulting in the \gls{lodcloud}, also referred to as the Web of Data, in which hundreds of diverse data sources enjoy varying degrees of semantic integration through links, with a variety of access points \citep{Bizer:2009:linked}.\footurln{See for example}{http://thedatahub.org}
%defined by  as ``a web of things in the world, described by data on the Web'' (p. 2).
%The more data is available as \gls{lod}, the more connections can be discovered between datasets, delivering rich and relevant results to end-users.

%Access points and repositories of \gls{lod} are maintained online.
%Recently, the potential of the paradigm has been highlighted by Google's Knowledge Graph, a large support tool that utilises Freebase, a \gls{lod} resource, to semantically enrich the search engine's results \citep{singhal:2012:googleknowgraphchicago}.

% in GIScience
\changed{As a large part of online data involves a spatial dimension, geographic entities and their semantics play a central role in the \gls{lodcloud}, facilitating the geospatial grounding of scientific and commercial data} \citep{hart:2013:lodgeobook,janowicz:2012:geospatiallinked}.
%Semantics is key to enable the usage, integration, and exploration of geographic data .
\changed{
The \gls{lod} paradigm is promising in the context of \gls{gir}, where existing techniques have shown limited effectiveness \citep{purves:2011:geographic}.
For example, the LOD-based search engine Wikipedia Faceted Search handled complex geospatial queries, e.g. `Which Rivers flow into the Rhine and are longer than 50 kilometers?' \citep{hahn:2010:facetedwikisearch}.}
% VGI
The emergence of the \gls{lod} infrastructure also has great potential for the dissemination of geographic data.
A prominent example is found in the British Ordnance Survey, which has embraced the paradigm and released some of its informational assets as \gls{lod}\footurl{http://data.ordnancesurvey.co.uk} \citep{goodwin:2008:geographical}.
%In parallel, \gls{vgi} is gaining credibility as a source of detailed information generated by non-expert users through crowdsourcing \citep{dodge:2013:mappingexp}. 
%Challenging traditional top-down cartographic engineering, \gls{osm} provides an open platform to build a world map, tapping its contributors' knowledge of their local geographic and social world \citep{Coast:2010:bestmapchicago}.
%To date, a gap between \gls{vgi} datasets and the \gls{lodcloud} exists, and constitutes a barrier to the integration and usage of the data.

% our work 
To enable the promising network effects in the \gls{lodcloud}, datasets need to be inter-connected through meaningful relationships.
Generating such semantic mappings automatically is therefore a crucial part of the \gls{lod} vision, enabling interoperability while preserving local semantic details.
In the \gls{lod} jargon, the process of linking a new dataset to existing ones is called `bootstrapping,' and is usually performed on semantic hubs such as \gls{dbp} \citep{mendes:2011:dbpspotlight}.
In this article, extending a preliminary study \citep{Ballatore:2013:groundinglod}, we focus on the bootstrapping of geographic vocabularies, utilising \gls{wn} as a \gls{lod} hub.

In this context, we first describe \gls{algo}, a generic technique to generate a semantic mapping between a given vocabulary and \gls{wn}, which we selected as a shared semantic ground because of its rich relations \citep{fellbaum:2010:wordnetsurvey}.
\removed{This mapping is not the goal in itself, but can support and enable a number of \gls{nlp} and \gls{ir} operations on geographic \gls{lod}.}
\changed{This semantic mapping is valuable because it can support and enable a number of \gls{nlp} and \gls{ir} operations on geographic \gls{lod}.
\gls{algo} is aimed at the underspecified vocabularies adopted in \glspl{gkb},} to increase their interoperability, and to enable the discovery of rich ontological relations such as part-whole (e.g. part-of relations) and subsumption (e.g. is-a relations), which are present in \gls{wn}.
%First, we describe a \gls{lod} resource, the \gls{osn}, which offers a machine-readable, structured, open conceptualisation of \gls{osm} semantics.
Second, we evaluate \gls{algo} on two real datasets containing primarily geographic information, the crowdsourced \gls{osn} and the lightweight \gls{gno} which provides a vocabulary to a large digital gazetteer.

%We extracted this semantic network from the \gls{oww} and other sources to provide a semantic support tool to interpret, search, and tap the vast vector dataset generated by \gls{osm} contributors.
%The \gls{osn} is currently integrated into the \gls{lodcloud}.

% article structure
The remainder of this article is organised as follows.
Section \ref{sec:relwork} reviews relevant work in the areas of \gls{lod} integration, open \glspl{gkb}, geo-semantics, and \gls{wn}.
This section also describes the \gls{osn} and the \gls{gno}, which are used in the evaluation. 
Section \ref{sec:mappingOsnToWn} describes and formalises \gls{algo}, a generic approach to semantic mapping onto \gls{wn}. 
Subsequently, we report on the evaluation of the approach, executed on a sample of terms from the \gls{osn} and the \gls{gno}, and compared with existing LOD mapping tools in Section \ref{sec:eval}.
Finally, conclusions and directions for future research are discussed in Section \ref{sec:concl}.

%=============================================
\section{Related work}
\label{sec:relwork}

%This section discusses related work relevant to the semantic alignment of geo-vocabularies, including \gls{lod} integration, \gls{wn}, the \gls{osn}, and the \gls{gn} gazetteer.

% ------------------------------------------------- %

The approach to \gls{lod} integration proposed in this article is inscribed in the \gls{sgw} research, in which identification of the same concepts and entities in heterogeneous data spaces through semantic similarity measures is considered to be a crucial enabler \citep{janowicz:2012:geospatiallinked}.
%To generate \gls{lod}, it is necessary to link the new entities to existing ones in the \gls{lodcloud}, a process often called `bootstrapping' \citep{mendes:2011:dbpspotlight}.
More generally, the automatic merging of different conceptual schemas is a time-honoured challenge in computer science, beginning well before the advent of the \gls{sw}.
Two datasets can be aligned at the schema level (e.g. matching the concept `river' in both ontologies), and at the instance level (e.g. connecting the Po River in both knowledge bases).
Logical reasoning, machine learning, and statistical analysis have been utilised to tackle the problem in the context of database schemas \citep{noy:2004:semintegration}.
Since 2005, the Ontology Alignment Evaluation Initiative (OAEI) has proposed benchmarks and performance metrics specifically tailored to the area of ontology alignment and integration \citep{euzenat:2011:oaei}.

Several approaches to generate a mapping have been devised, both from an intensional and an extensional viewpoint.
\emph{Terminological} methods rely on simple string matching between the terms, while \emph{semantic} methods compare the representation of terms in formal semantic models.
Furthermore, semantic methods can observe the terms from multiple angles: \emph{internal} methods observe aspects of the terms in isolation, such as the attribute ranges.
By contrast, \emph{external} methods analyse the relational structure of the ontologies, comparing the position of the terms relative to the other terms.
Finally, \emph{extensional} methods perform the alignment based on distributional properties of term instances.
As covered in the next section, these approaches are utilised in actual information integration software tools.

\subsection{LOD integration frameworks}
\label{sec:relwork_opendataint}

% survey of existing tools
To perform the integration of \gls{lod} datasets stored in RDF format, a number of frameworks have been developed. 
The RDF-AI tool aims at the integration of RDF datasets \citep{scharffe:2009:rdfai}.
The matching is performed by computing the semantic similarity of two given entities, based on a user-provided set of salient properties (e.g. the title and year of a musical work, the author and title of a book, etc.).
The semantic similarity can be computed either with fuzzy string matching based on the sequence integration algorithm, or by comparing synonyms in \gls{wn}.
Subsequently RDF-AI uses the matching pairs either to fuse two datasets into one, or to generate a list of matching entities.

Along similar lines, \cite{volz:2009:silklinkdiscovery} developed \emph{Silk Link Discovery Framework}, which aims at establishing relations between entities in different data sources.
A number of strategies can be used to match properties, based on simple string similarity measures.
The user can specify what properties should be compared and with which similarity metric, and can specify the thresholds above which the relations should be established or should be manually verified.
For example, in a given context, all pairs with similarity equal to or greater than $0.9$ might be linked automatically, while pairs with similarity greater than $0.6$ but smaller than $0.9$ should be checked manually.  
Such heuristics can be defined in the Link Specification Language (Silk-LSL).
More recently, \cite{isele:2012:expressivelinkage} extended Silk with the \emph{GenLink} algorithm, which extracts rules from valid links using supervised machine learning.

Scalability issues affect these tools, which often are crippled by the enormous complexity of the brute-force comparison of large datasets.
To overcome this issue, \cite{ngomo:2011:limeslargelink} developed the \emph{LInk discovery framework for MEtric Spaces} (LIMES).
This framework performs operations logically equivalent to those of Silk, but relies on the concept of triangle inequality in metric spaces to compute pessimistic estimates of instance similarities.
Based on these approximations, LIMES can exclude a large number of entity pairs that cannot satisfy the user-defined matching conditions.
The actual similarities of the remaining pairs are then computed and the matching instances are returned, without losing recall.
While these frameworks are useful in the context of a generic matching between entities in \gls{lod} datasets, they do not perform well in the case of \gls{wn}, as discussed in Section \ref{sec:evalTools}.

% LINKED DATA

%Linked Data is about using the Web to connect related data that wasn't previously linked, or using the Web to lower the barriers to linking data currently linked using other methods. More specifically, Wikipedia defines Linked Data as ``a term used to describe a recommended best practice for exposing, sharing, and connecting pieces of data, information, and knowledge on the Semantic Web using URIs and RDF.''\footurl{http://linkeddata.org}

%The inter-connected network of diverse and heterogenous datasets is dubbed the `\gls{lod} cloud.'\footurl{http://richard.cyganiak.de/2007/10/lod}

%The Open Knowledge Foundation .\footurl{http://okfn.org} 

% ------------------------------------------------- %
\subsection{WordNet as a semantic hub}

Since the early 1990s, \Gls{wn} has been a valuable semantic resource for many applications in \gls{nlp} and artificial intelligence \citep{fellbaum:1998:wordnet,fellbaum:2010:wordnetsurvey}.
The core element of \Gls{wn} is the `synset,' a concept that aggregates a set of synonymous words, called `word senses.'
For example, the geographic concept `stream' is represented in \gls{wn} by synset $\{$\semuri{stream,watercourse}$\}$.
This synset contains two word senses, \semuri{stream\#n\#1} and \semuri{watercourse\#n\#1}, with the notation \semuri{word\#part-of-speech\#word-sense-number}.
The word `stream' appears in five different synsets, capturing its high polysemy.
Synsets are connected through several semantic relations, such as \emph{similarTo}, \emph{partMeronymOf}, \emph{adjectivePertainsTo}, \emph{causes}, \emph{antonymOf}, and \emph{entails}.\footnote{See \url{http://www.w3.org/2006/03/wn/wn20/schemas/wnfull.rdfs} for the complete list.}
Two versions of \gls{wn}, 2.0 and 3.0, are currently linked in the \gls{lodcloud}.\footnote{\url{http://www.w3.org/2006/03/wn/wn20} and  \url{http://semanticweb.cs.vu.nl/lod/wn30}}

\Gls{wn} has found particular success in the areas of \gls{wsd} and semantic similarity \citep{navigli:2009:word,Ballatore:2012:jury}.
Different components of the network have been exploited to model the semantic similarity of its synsets, tapping its deep taxonomy, and the word definitions, called `glosses' \cite[e.g.][]{ramage:2009:random}. 
Although the semantic network was not designed for this purpose, it has been frequently used as a general-purpose semantic ground, for example to discover semantic connections in unstructured data \citep{lin:2009:folk2onto}.
% limitations of WordNet
The limitations of \gls{wn} have been thoroughly discussed.
Being a top-down, expert-controlled resource, its lexical coverage is bound to be lower than that of crowdsourced alternatives, such as \gls{dbp}.
Furthermore, the upper part of its taxonomical structure has been critised as ontologically unsound, prompting a substantial re-design and refinement, following state-of-the-art ontological theories \citep{gangemi:2003:wordnetdolce}.

% derived projects 
A large number of projects provide \gls{wn}-like semantic networks in languages other than English.\footurln{See the list at}{http://www.globalwordnet.org/gwa/wordnet_table.html}
%GeoWordNet, for example, aggregates WordNet synsets with the open gazetteer GeoNames \citep{giunchiglia:2010:geowordnet}.
To date, none of the numerous alternative semantic resources has yet managed to dethrone \gls{wn} from its leading position as general-purpose semantic ground.
In the context of the \gls{lodcloud}, \gls{wn} has been used as a high-quality primary semantic source in many projects inter-linked with \gls{dbp}, the largest hub of the \gls{lodcloud} \citep{Ballatore:2012:survey}.
%More recently, a \gls{sw} version of \gls{wn} 3.0 has been released, with mappings to version 2.0.
%These resources are .
Although \gls{dbp} has considerably larger coverage than \gls{wn}, its ontological structure is lighter, and provides fewer semantic relations.
For this reason, we argue that \gls{wn} could complement \gls{dbp} as a central resource in the \gls{lodcloud}.
Using \gls{wn} as an imperfect, and yet rich semantic ground, it is possible to integrate geo-vocabularies, such as the \gls{osn} and the \gls{gno}, described in the next sections.

% ------------------------------------------------- %
\subsection{The \protect\gls{osn}}
\label{sec:osn}

\glsreset{vgi}
\glsreset{osm}
\glsreset{skos}

\removed{Crowdsourced geographic information, also referred to as} \Gls{vgi} is playing an increasingly important role in the \gls{lodcloud}.
From its foundation in 2004, \gls{osm} has established itself as the most ambitious \gls{vgi} project.
The \gls{osm} conceptualisation emerges from semantic negotiations within the contributors' community, reaching consensus around the intended meaning and usage of `tags,' i.e. terms describing geographic entities. % and, to a lesser extent, processes.
%This semi-structured folksonomy allows contributors to create any new term to describe the objects that they find worth mapping, using a dedicated wiki website to define, discuss, and use terms \citep{vander:2007:folksonomychicago}.
%For example, terms \texttt{landuse=forest} and \texttt{natural=wood} are used to label areas covered by trees, forestry and woodland.\footurl{http://wiki.openstreetmap.org/wiki/Tag:landuse=forest}
\changed{This radically open approach to geo-semantics was adopted by the project's creators on the assumption that an all-encompassing geographical ontology is an unrealistic endeavour, and that a bottom-up negotiation allows for more experimentation, and attracts non-expert contributors.}
%On the other hand, a traditional conceptual modelling approach would have arguably raised a barrier for contributors. %, or, as project founder Steve \citet{Coast:2010:bestmapchicago} put it, ``to dictate [terms] as in a top-down ontology would have been nuts.''
The downside of the adoption of a semi-structured folksonomy is, predictably, wide variability and ambiguity in the terms' interpretation, proliferation of near-synonym terms, and lack of explicit semantic relations \citep{Ballatore:2011:semantically}.
The \gls{osn} is interlinked with \gls{lgd} and \gls{dbp} \citep{Auer:2009:linkedgeodata}.
Using \gls{algo}, described in Section \ref{sec:mappingOsnToWn}, the network has also been linked to \gls{wn}.
%, resulting in a `spatially rich and semantically poor' dataset .

% OSM semantic projects
%In recent years, efforts have been undertaken to strengthen the thin semantic ground on which \gls{osm} rests.
%A valuable resource in the \gls{lodcloud} is the \gls{lgd} project \citep{Auer:2009:linkedgeodata}.
%\gls{lgd} provides a \gls{sw} version of the \gls{osm} dataset, released as \gls{lod}.
%Despite the increased accessibility to individual map features, the semantics of \gls{osm} terms contained in the \gls{oww} is not included in the \gls{lgd}.
%From a different perspective, OSMOnto provides an ontology that structures the most common \gls{osm} terms.\footurl{http://wiki.openstreetmap.org/wiki/OSMonto}

%Furthermore, \citet{baglatzi:2012:semantifying} devised an approach to grounding the \gls{osm} folksonomy on the DOLCE upper-level ontology \citep{gangemi:2002:dolce}.
%Acknowledging the extreme difficulty in implementing such semantic mapping in an automatic way, they designed a \gls{gwap} to crowdsource a human-quality mapping.
%In our previous work, we devised an initial semantic integration between \gls{osm} and \gls{dbp}, geared towards exploratory navigation of Web maps \citep{Ballatore:2011:semantically}.
To provide a knowledge-based support tool for \gls{osm}, we extracted the \gls{osn}, a semantic artefact containing the conceptualisation of \gls{osm} tags, providing a machine-readable structure that can support the automatic manipulation of \gls{osm} features in data mining, \gls{gir}, and information integration \citep{Ballatore:2013:groundinglod}.\footurl{http://wiki.openstreetmap.org/wiki/OSMSemanticNetwork} 
% \citep{Ballatore:2012:geographic,Ballatore:2013:groundinglod}.
%Integrating the \gls{osn} as \gls{lod}, grounded with \gls{wn}.
%The semantic network was utilised to compute the semantic similarity of \gls{osm} terms using link-based measures \citep{}.
%The \gls{osn} is
The network was initially developed offline to compute the semantic similarity of tags \citep{Ballatore:2012:geographic}, and is published in the \gls{lodcloud}.\footurl{http://datahub.io/dataset/osm-semantic-network}
% why SKOS
The \gls{osn} is organised as a \gls{skos} vocabulary \citep{miles:2005:skos}.
\Gls{skos} is a semantic formal language designed to allow the publication and sharing of technical vocabularies, taxonomies, and classification systems.
In a \gls{skos} scheme, the main semantic unit is the \semuri{skos:Concept}.
A \emph{concept} is a term that can be defined using lexical definitions and linked to other concepts through semantic relations.

The semantic relations in SKOS are explicitly left as generic as possible.
Concepts can be more general or specific than other concepts (\semuri{skos:broader} and \semuri{skos:narrower}), and can be semantically related (\semuri{skos:related}).
A concept is described by a preferred short lexical label (\semuri{skos:prefLabel}), and can have $n$ alternative labels (\semuri{skos:altLabel}).
A more extensive and unique definition can be given to a concept in a given language (\semuri{skos:definition}).
Hence, each term defined in the network corresponds to a SKOS concept.
For example, the OSM tag \osmtag{waterway=river} corresponds to the term \semuri{osnt:k:water\-way/\-v:river}.\footurl{http://spatial.ucd.ie/lod/osn/term/k:waterway/v:river}
The quality of the \gls{skos} vocabulary was assessed based on the criteria outlined by \citet{suominen:2012:improving}.
Another example of a SKOS-based vocabulary is the \gls{gno}, described in the next section.

\subsection{The GeoNames ontology}
\label{sec:geonames}

The \gls{gn} project is an open digital gazetteer combining a variety of data sources, representing the location of about 8 million unique features.\footurl{http://www.geonames.org}
Thanks to its impressive coverage, this gazetteer is widely used in geospatial applications, and constitutes a densely linked resource in the \gls{lodcloud}.
The geographic features contained in \gls{gn} are classified using a simple hierarchical tree, in which 9 Feature Classes (e.g. \emph{Populated places}) contain more specific 690 Feature Codes (e.g. \emph{religious populated places}).
Although this artefact is a lightweight SKOS vocabulary with little formal ontological content, it is referred to as the \gls{gno}, and has reached version 3.1.

The peculiarities and issues found in the \gls{gno} have been discussed by \cite{giunchiglia:2010:geowordnet}, who integrated it manually with \gls{wn} to generate \gls{gwn}, a geographically enhanced version of \gls{wn}.
Although this integration provides indeed a useful resource, our contention is that automated interlinking should be preferred to the manual semantic merging applied in \gls{gwn}.
Even if automated semantic bootstrapping is unlikely to equal manual mapping in terms of quality, it provides a sustainable way to include new resources in the \gls{lodcloud}, without increasing the fragmentation of existing resources into multiple versions and preserving the structure of each resource and their local semantics.

%The \gls{gno} is utilised to evaluate \gls{algo}, 

% Semantic similarity in GIR
%To be interpreted by humans, the geographic information represented in viewports has to convey some intelligible meaning.
%The semantics of geographic data has been discussed extensively by , who points out the difficulties of grounding meaning in symbolic systems.

%In order to compare, classify, index and cluster geographic objects by their semantics, several analytical approaches have been devised.
%\citet{schwering:2008:approaches} provides a thourough survey and classification of the the similarity measures for geographic data.

%\gls{algo} bears strong resemblance with \gls{dbp} Spotlight, a semantic tagger that identifies \gls{dbp} resources in raw text \citep{mendes:2011:dbpspotlight}.
% ------------------------------------------------- %
% link to next section 

%The next section describes our contribution to the area of \gls{vgi} integration into the \gls{lodcloud}.
In this sense, whilst \gls{gwn} is the result of a merging process, resulting in a new resource, \gls{algo} provides an automatic mapping technique between a given vocabulary and \gls{wn}.
%In this sense, \gls{algo} is to provide an automatic mapping technique between a given vocabulary and \gls{wn}, whilst \gls{gwn} is the result of a merging process, resulting in a new resource.
To the best of our knowledge, a semantic mapping technique between a vocabulary and \gls{wn}, geared towards the `bootstrapping' of the vocabulary in the \gls{lodcloud}, has not been devised, and 
\gls{algo} has precisely the purpose of filling this specific gap.
In this sense, it is not a general-purpose ontology mapping technique.
As described in the next section, \gls{algo} performs the semantic mapping between a vocabulary term and a specific \gls{wn} word sense both from an intensional (i.e. lexical overlap between the lexical definitions) and an extensional perspective (i.e. the usage frequency).
%It should not come as a surprise that expressing identity constitutes one of the key problems of the \gls{lodcloud}, which  dubbed `the identity crisis of Linked Data.'

%=============================================
\section{\protect\gls{algo}, a semantic mapping algorithm}
\label{sec:mappingOsnToWn}
%=============================================

\glsreset{lod}

\begin{table}[t]
\centering
\begin{tabular}{cP{1em}P{28em}}
		\hline
		Symbol & & Description\\
	    \hline
	    $V$ &  & Vocabulary, i.e. set of terms $t$. E.g. the \gls{gno}\\
	    $t$ &  & Generic term $\in V$. E.g. \semuri{osnt:k:waterway} \\
	    %$t_k$ &  & Key term in $\in V$. E.g. \semuri{osnt:k:waterway} \\
	    %$t_v$ &  & Tag term in $\in V$. E.g. \semuri{osnt:k:waterway/v:river} \\
	    $\Theta$ &  & Salient taxonomy extracted from \gls{wn}. \\
	    $W$ &  & \gls{wn}, i.e. a set of synsets. \\
	    $s$ &  & \gls{wn} synset, $s \in W$. E.g. \semuri{wn:river-noun-1}\\
	    $ws$ &  & Word sense in synset $s$. E.g. \semuri{wn:wordsense-river-noun-1}\\
	    $C_t$ &  & Candidate synsets $s \in W$ for term $t$\\
	    $ol(t,s)$ & & Overlap between definitions of term $t$ and synset $s$. $ol \geq 0$\\
	    $f(ws)$ & & Usage frequency of $ws \in s$. $f \geq 0$ \\
	    $ol_{min}$ &  & Minimum lexical overlap between terms. \\
	    $f_{min}$ &  & Minimum frequency of word sense in \gls{wn}. \\
	    $\sigma(s,ws,t)$ &  & Salience score for candidate $s$ and $ws$ for term $t$. \\
	    $M(V, W)$ &  & Set of semantic mappings $m$ between vocabulary $V$ and $W$ \\
	    $m$ &  & Semantic mapping $<t,r,s>$ between term $t \in V$ and synset $s \in W$, with relation $r$\\
	    $r$ &  & Relation that defines the nature of the semantic mapping $m$: exact, close, or related (see Section \ref{sec:mappingDetails}) \\
	    %$M_h$ &  & Human-generated semantic mapping between $V$ and $W$. \\
	    %$P_M$ &  & Precision of semantic mapping $M$ with respect to $M_h$\\
	    %$R_M$ &  & Recall of semantic mapping $M$ with respect to $M_h$\\
	    %$F_{\beta M}$ &  & $F$-measure, biased towards precision ($\beta = .5$) \\
	    \hline 
\end{tabular}
\caption{Notations}
\label{table:notations}
\end{table}

\begin{table}[t] 
% VERY GOOD TRICK TO JUSTIFY TABLE! (package: graphics)
\resizebox{\columnwidth}{!}{%
\begin{tabular}{lll}
		\hline
		Abbr. & Description & URI\\
	    \hline

	    %$osnt$ & OSM term & \shortu{http://spatial.ucd.ie/lod/osn/term/} \\
	    %$osnt$ & $-$ key & \shortu{http://spatial.ucd.ie/lod/osn/term/k:<key>} \\
	    
	    %$osnp$ & $-$ property & \shortu{http://spatial.ucd.ie/lod/osn/property/} \\
	    
	    %\hline
	    %$owl$ & OWL & \shortu{http://www.w3.org/2002/07/owl#} \\
	    $rdfs$ & RDF schema & \shortu{http://www.w3.org/2000/01/rdf-schema#} \\
	    $skos$ & SKOS & \shortu{http://www.w3.org/2004/02/skos/core#} \\
	    $wn$ & \gls{wn} synset & \shortu{http://www.w3.org/2006/03/wn/wn20/instances/synset-} \\
	    $ws$ & $-$ word sense & \shortu{http://www.w3.org/2006/03/wn/wn20/instances/wordsense-} \\
	    $wns$ & $-$ schema & \shortu{http://www.w3.org/2006/03/wn/wn20/schema/} \\
	   	$osn$ & \gls{osn} & \shortu{http://spatial.ucd.ie/lod/osn/} \\
	    $osnt$ & $-$ tag & \shortu{http://spatial.ucd.ie/lod/osn/term/k:<key>/v:<value>} \\
	    $osnpt$~~ & $-$ proposed term ~~& \shortu{http://spatial.ucd.ie/lod/osn/proposed_term/} \\
	    $gno$ & GeoNames ontology & \shortu{http://www.geonames.org/ontology#} \\
	    $lgdo$ & \gls{lgd} & \shortu{http://linkedgeodata.org/ontology/} \\
	    %$dc$ & Dublin Core & \shortu{http://purl.org/dc/elements/1.1/}\\
	    %$lgdo$ &  & \semuri{} \\
	    \hline 
\end{tabular}
}
\caption{XML namespaces}
\label{table:ns}
\end{table}

To increase integration and interoperability of \gls{lod} at the schema level, we propose to utilise the lexical database \gls{wn} as a semantic hub.
For this purpose, this section describes \gls{algo}, an algorithm devised to generate a semantic mapping between a given vocabulary and \gls{wn}.
The algorithm generates a semantic mapping between a given vocabulary $V$ containing a set of terms (e.g. a \gls{skos} vocabulary), and \gls{wn} synsets that are semantically similar.
The issue tackled by \gls{algo} is inscribed within the open problem of \gls{wsd}, i.e. distinguishing when the word `bank' refers to  a financial institution or to the terrain alongside a river \citep{navigli:2009:word}.
The similarity notwithstanding, the constraints in which \gls{algo} operates make the \changed{integration} considerably simpler than open \gls{wsd} on raw text.

The \gls{algo} approach is primarily aimed at the schema level typical of vocabularies, and not at the instance level, and combines intensional and extensional aspects to identify salient synsets in \gls{wn}.
% One of key tenets of \gls{lod} is the importance of inter-linking.
% When heterogeneous datasets are semantically linked, new connections are enabled, enabling cross-data analysis and supporting knowledge extraction.
% For this reason, we adopt the \gls{lod} paradigm and we link \gls{osn} to other datasets in the \gls{lodcloud}.
Although this article focuses on geo-vocabularies, \gls{algo} can be used to map any vocabulary into \gls{wn}.
The notations used in the remainder of this article are reported in Table \ref{table:notations}.
For the sake of brevity, the namespaces are summarised in Table \ref{table:ns}.
Section \ref{sec:mappingDetails} defines the nature and scope of the semantic mapping for which \gls{algo} is designed.
The detailed workings of \gls{algo} are subsequently described in Section \ref{sec:algo}. 

\subsection{Mapping relations}
\label{sec:mappingDetails}

A semantic mapping $m$ between term $t \in V$ and synset $s \in W$ has the form $<t,r,s>$.
Given the aim of \gls{skos} to provide a Web and collaborative platform for vocabularies, the language provides semantic relations to connect concepts to equivalent, similar or related concepts in other vocabularies.
Such relations are called \emph{mapping properties}.\footurl{http://www.w3.org/TR/skos-reference/#mapping}
A concept can engage in an identity relation with a concept in another schema (\semuri{skos:exactMatch}), can be very similar (\semuri{skos:closeMatch}), or can be only loosely related to it (\semuri{skos:relatedMatch}). 
%In the \gls{osn}, we adopt a fine-grained approach.
%From a conceptual modelling perspective, the issue of identity have been thoroughly studied .
In the context of \gls{algo}, we adopt three \gls{skos} symmetric mapping relations $r$:

\begin{description}
	\item[Related] (\semuri{skos:relatedMatch}): General semantic relatedness (e.g. \semuri{osnt:\-k:power/\-v:station} and \semuri{wn:electricity-noun-1}); 
	\item[Close] (\semuri{skos:closeMatch}): Highly similar terms which originated from different information communities (e.g. \semuri{osnt:k:wood} and \semuri{wn:forest-noun-2});
	\item[Exact] (\semuri{skos:exactMatch}): Terms that originated from the same information community, but expressed in different vocabularies (e.g. \semuri{osnt:k:amenity/\-v:uni\-ver\-sity} and \semuri{lgdo:Uni\-ver\-sity}).  We consider this mapping to be logically equivalent to \semuri{owl:sameAs}.
\end{description}

\begin{figure}[t]
  \centering
  % figure generated from file: 
  %\includegraphics[width=34em]{figures/osn_diagram_lod.pdf}
    \includegraphics[width=35em]{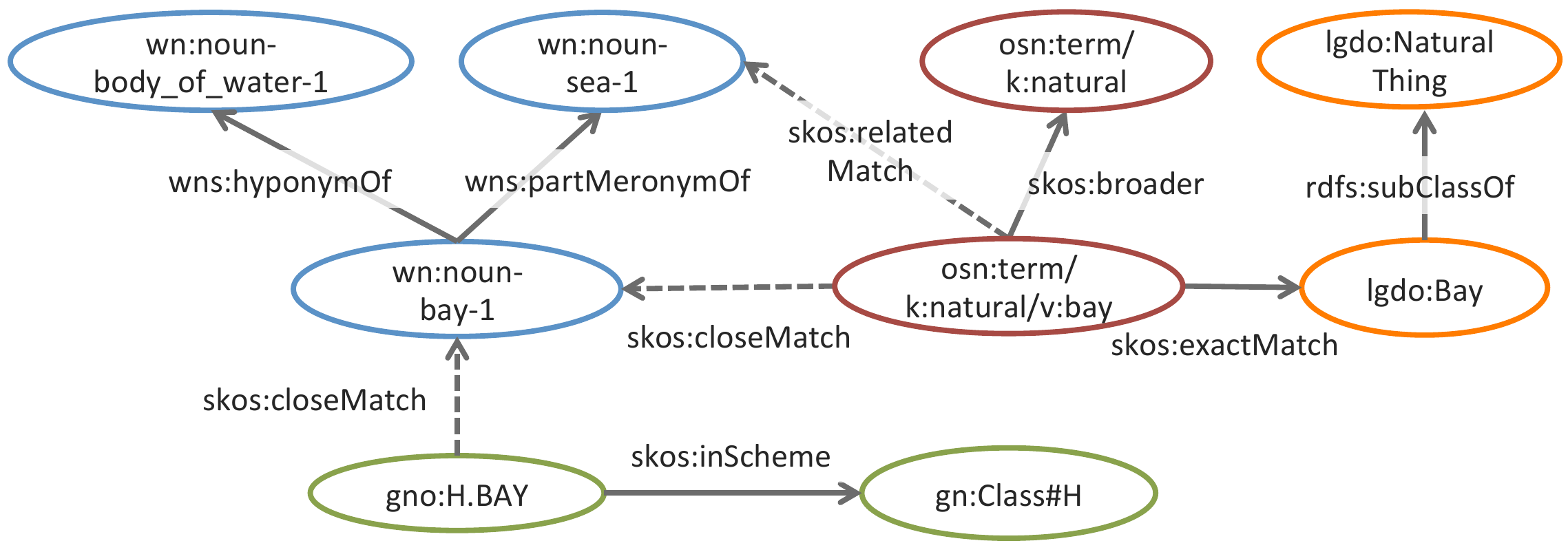}
  \caption{Fragments of entities representing geographic concept `bay' and their mappings in \gls{wn} \changed{(\emph{wn}), \gls{lgd} (\emph{lgdo}), the \gls{osn} (\emph{osn}), and the \gls{gno} (\emph{gno})}. Dotted relations are generated by \gls{algo}.}
  \label{fig:fragmentmapping}
\end{figure}

\noindent Through these relations, it is possible to establish a mapping $m=<t,r,s>$ between the vocabulary $V$ and the \gls{wn} synsets $W$.
\changed{We define the validity of a mapping in terms of its semantic coherence (is the mapping's semantics clear to a human observer?) and completeness (does the mapping include all the possible coherent relationships?).}
Figure \ref{fig:fragmentmapping} shows a fragment of a possible valid mapping of the geographic term `bay' between the \gls{gno}, the \gls{osn}, \gls{lgd}, and \gls{wn}.
To further illustrate the difficulties of the semantic mapping with \gls{wn}, the definition of \semuri{wn:bay-noun-1} is ``an indentation of a shoreline larger than a cove but smaller than a gulf,'' while \semuri{wn:bay-noun-2} is defined as ``the sound of a hound on the scent,'' an alternative and semantically unrelated meaning.
The \gls{osm} term \semuri{osnt:k:natural/v:bay} is defined as a ``a large body of water partially enclosed by land but with a wide mouth.''
The following list shows possible correct and incorrect mappings between these terms:

\begin{enumerate}[label=(\alph*)]
  %\item $<$\semuri{osnt:k:amenity/v:restaurant} \emph{exact} \semuri{lgdo:Restaurant}$>$  (correct)
%  \item $<$\semuri{osnt:k:amenity/v:restaurant} \emph{close} \semuri{wn:restaurant-noun-1}$>$  (correct)
%  \item $<$\semuri{osnpt:\-Art\_gallery} \emph{close} \semuri{wn:gallery-noun-3}$>$  (correct)
%  \item $<$\semuri{osnpt:\-Art\_gallery} \emph{related} \semuri{wn:art-noun-1}$>$  (correct)
%  \item $<$\semuri{osnpt:\-Art\_galle\-ry} \emph{close} \semuri{wn:gallery-noun-1}$>$  (incorrect, wrong word sense)
%  \item $<$\semuri{osnpt:\-Art\_gallery} \emph{close} \semuri{wn:art-noun-1}$>$  (incorrect, wrong relation)
  
  \item $<$\semuri{osnt:k:natural/v:bay} \emph{close} \semuri{wn:bay-noun-1}$>$  (correct)
  \item $<$\semuri{osnt:k:natural/v:bay} \emph{related} \semuri{wn:sea-noun-1}$>$  (correct)
  \item $<$\semuri{osnt:k:natural/v:bay} \emph{related} \semuri{wn:bay-noun-1}$>$  (incorrect)
  \item $<$\semuri{osnt:k:natural/v:bay} \emph{related} \semuri{wn:bay-noun-2}$>$  (incorrect)
  \item $<$\semuri{osnt:k:natural/v:bay} \emph{close} \semuri{wn:sea-noun-1}$>$  (incorrect)

  %\item TODO: replace with examples from \gls{gno}
\end{enumerate}

\noindent Case (e) should considered incorrect because the synset `sea' is only \emph{related} to `bay,' and does not constitute a close match.
In some situations, the distinction between \emph{close} and \emph{related}, and \emph{close} and \emph{exact}, is more nuanced, and both cases can be considered correct.
% 

%\gls{dbp} Spotlight is a semantic tagger \citep{mendes:2011:dbpspotlight}.

\subsection{Algorithm}
\label{sec:algo}

\gls{algo} generates a mapping $M$ between a given vocabulary $V$ and the set of \gls{wn} synsets $W$.
Given a term $t \in V$, \gls{algo} utilises a lexical matching function on the words contained in the lexical definition of $t$, taking compound words into account (e.g. `swimming pool'), and then splitting them if not defined directly in \gls{wn} (e.g. `swimming' and `pool').
%Given a term $t \in V$, \gls{algo} utilises a lexical matching function on the words contained in $t$, taking compound words into account (e.g. `swimming pool'), and then splitting them if not defined in \gls{wn} (e.g. `swimming' and `pool').
If the set of matching wordsenses $ws$ is not empty, the algorithm relies on three indicators of semantic salience: 

\begin{description}
  \item[Word sense frequency $f$:] The usage frequency $f$ of a \gls{wn} word sense is correlated with its semantic salience. In the context of a shared vocabulary, common word senses are more likely to be correct than uncommon word senses.  For example, for $t=$`field', \semuri{ws:field-noun-1} (``a piece of land cleared of trees and usually enclosed'') has a usage frequency $f=49$, whilst \semuri{ws:field-noun-12} (``all of the horses in a particular horse race'') has $f=1$. Indeed, this assumption can be false in the context of open text.  
  
  \item[Lexical overlap $ol$:] Similar terms tend to be defined using the same words.  The lexical overlap $ol$ is the number of word shared by the lexical definitions of two terms.  Terms showing high lexical overlap are more likely to be salient than terms that do not show overlap.  The overlap is considered after the removal of stopwords, and lemmatisation, excluding the term that is being defined.  For example, the overlap between the definitions of term $t$ (``A river is a body of water'') and \semuri{wn:river-noun-1} (``Rivers are natural streams of water'') is equal to 1.
  
  \item[Salient taxonomy $\Theta$:] If a vocabulary is domain specific, the mapping can be restricted to a salient taxonomy $\Theta$, i.e. a subset of \gls{wn}.
  Salient word senses tend to engage in semantic relations with salient synsets.
  Looking at the noun taxonomy of \gls{wn}, it is possible to select high-level synsets that are salient to the vocabulary's domain.
  If the candidate synsets engage in some relation with such salient taxonomical roots, they are more likely to be valid than synsets that do not.  
  For example, let us choose \semuri{wn:artifact-noun-1} as a salient root, and `shelter' as $t$.
  It is possible to infer that \semuri{ws:shelter-noun-2} (``protective covering that provides protection from the weather'') is related to the salient root through a path of transitive subsumption relations (\semuri{wns:hyponymOf}), while \semuri{ws:shelter-noun-4} (``a way of organizing business to reduce the taxes it must pay on current earnings'') is not.
%   \item Extract salient taxonomy $\Theta$ from \gls{wn};
%   \item For each term $t \in V$ to be mapped, retrieve set of candidate synsets $W_t$;
%   \item Exclude candidates that do not belong to taxonomy $\Theta$;
%   \item Pre-process synset and term definitions ($POS$ tagging and stopwords filter);
%   \item Compute salience indicators $ol(t,s)$, and $f(s)$;
%   \item Exclude candidates that do not match criteria $ol_{min}$ and $f_{min}$;
%   \item Rank candidate $W_t$ based on salience indicators (term overlap $ol$ and frequency $f$);
%   \item Select best candidate(s), and assign mapping confidence;
%   \item Add mapping $m$ to set $M$.
\end{description}

\noindent Formally, we define $t$ as the input term, $C_t$ as the set of candidates for term $t$, $ws$ as the candidate word sense, $s$ as the corresponding synset, and $\Theta$ as a manually selected salient taxonomy.
The non-negative $\theta$ is set to $1$ if $s \in \Theta$, and $0$ otherwise.
The salience of the three indicators are captured in a normalised score $\sigma$ as follows:

\begin{align}
\sigma(t, ws, s) = \frac{2|C_t| - rank(f(ws)) - rank(ol(t,s)) + \theta}{2|C_t|-1} \\  
\sigma \in [0,1],~ rank \in [1,|C_t|]\nonumber\\
 \theta=1 ~ if(s \in \Theta), \theta = 0 ~ otherwise\nonumber
\end{align}

\noindent The salience score $\sigma$ captures the semantic similarity between term $t$ and the synset $s$, through the word sense $ws$, relative to the set of candidates $C_t$.
The ranking function $rank$ is applied on the set $C_t$, and returns an integer between $1$ and $|C_t|$.
The score falls in the interval $[0,1]$, where $0$ indicates no salience, and $1$ maximum salience. 
For example, given a $C_t$ with three candidates, if $ws$ and $s$ have the highest frequency ($rank(f) = 1$), the second highest overlap ($rank(ol)=2$), and $s$ belongs to the salient taxonomy $\Theta$ ($\theta = 1$), then $\sigma = .8$.

These three indicators are combined to select valid mappings both from the term itself $t$, and from the term's lexical definition, which can contain useful pointers to relevant terms (e.g. the definition of term `power station' contains `electricity').
In order to provide more leverage, the algorithm filters out candidates based on a minimum frequency ($f_{min}$), a minimum overlap ($ol_{min}$), and a manually selected salient taxonomy ($\Theta$).
The detailed workings of the algorithm and functions are outlined in Algorithm \ref{algo:algo}.
In the next section, \gls{algo} is evaluated on two real-world datasets, i.e. the \gls{osn} and the \gls{gno}.

% crawler algorithm
% http://janela.lirmm.fr/~fiorio/AlgorithmSty/algorithm2e.pdf
%\IncMargin{1em}
\begin{algorithm}[p]
\SetKwInOut{Input}{input}\SetKwInOut{Output}{output}
\Input{vocabulary $V$,
set of synsets $W$,
min overlap $ol_{min}$,
min word sense frequency $f_{min}$,
salient taxonomy $\Theta$}%\\$\sigma$ = maximum length for key/value strings (e.g. 30 characters)}
%\Output{directed graph $\mathbf{G} = (V,E)$}
\Output{Set $M$ of semantic mappings $m = <t,r,s>$ }
\BlankLine
%$V \leftarrow \emptyset$\\
$M \leftarrow \emptyset$\\
\ForEach{term $t \in V$}{
	{ $m \leftarrow$ findSemanticMapping($t,W$);\\}
	{ add $m$ to $M$;\\}
	{ extract terms from lexical definition of $t$ to set $D_t$;\\}
	\ForEach{term $d \in D_t$}{
		{ $m_d \leftarrow$ findSemanticMapping($d,W$)\\}
		{ set `related' as $r$;\\}
		{ add $m_d$ to $M$;\\}
	}
}
return $M$.\\
% \SetKwFunction{KwMatch}{LexicalMatch}
% \Func{
% \KwMatch{piach}{
% {some stuff\\}
% {some other stuff\\}
% }}
\caption{\protect\gls{algo}($V,W,ol_{min},f_{min},\Theta$)}
\label{algo:algo}
\end{algorithm}

\begin{function}[p]
\caption{findSemanticMapping($t, W$)}
$C_t \leftarrow \emptyset$\\
\ForEach{$ws \in W$}{
		{ find set of matching word senses $ws \in W$ with lexicalMatch($ws,t$);\\ }
		{  find synset $s$ corresponding to $ws$ in \protect\gls{wn}; \\ }
		{  if $s \notin \Theta$, skip $ws$; \\ }
		{  fetch word sense frequency $f(ws)$ from \protect\gls{wn};\\ }
		{  if $f(s) < f_{min}$, skip $ws$;\\ }
		{  compute lexical overlap between definitions $ol(s,t)$;\\ }
		{  if $ol(s,t) < ol_{min}$, skip $ws$;\\ }
		{  $s$ and $ws$ are a valid candidate, add pair $<s,ws>$ to candidate set $C_t$; \\ }
	}
	\ForEach{ $<s,ws>\in C_t$ }{
		{  compute salience score $\sigma(s,ws,t)$;\\ }
	}
	{ select best candidate $s_b \in C_t$ having $max(\sigma(s,ws,t))$; \\ }
	\eIf{lexicalMatch($ws,t$) is `complete' $\wedge ~max(ol(s,t)) \wedge max(f(ws))$}
		{select `close' as $r$ }
		{select `related' as $r$}
	%{ select best mapping relation $r_b$: if $t$ label is equal to $ws$; \\ }
	{ generate mapping $m=<t,r,s_b>$ and return it. \\ }
\end{function}

\begin{function}[ht]
\caption{lexicalMatch($ws, t$)}
	{ \If{$ws$ \emph{is contained in} $t$}{return `partial'; }
	  \If{$ws$ \emph{is equal to} $t$}{return `complete'; }
	   {return `no match'. \\} }
\end{function}

%=============================================
\section{Evaluation}
\label{sec:eval}
%=============================================
% Code in file wn_mapping_analysis.R

This section describes an experimental evaluation of \gls{algo}, our semantic mapping technique, outlined in Section \ref{sec:mappingOsnToWn}, which extends an initial exploration of the algorithm \citep{Ballatore:2013:groundinglod}.
%The technique is utilised to obtain a semantic mapping between \gls{wn} and terms defined by the .
We generated two evaluation datasets $M_h$ by selecting random samples of terms from the \gls{osn} and the \gls{gno} (Section \ref{sec:evalDs}).
To measure the performance of the algorithm, we defined performance measures (precision, recall, and an $F$-measure) that compare the machine-generated mapping $M$ with the human mapping $M_h$ (Section \ref{sec:evalMeasures}).
In order to compare \gls{algo} with existing tools, preliminary experiments were conducted on the mapping framework LIMES (Section \ref{sec:evalTools}).
%This evaluation extends the preliminary study on \gls{algo} \citep{Ballatore:2013:groundinglod} in two ways.
Finally, an experiment on a number of parameter combinations was executed on both datasets (Section \ref{sec:salientExpSetup}), and the performance of \gls{algo} is analysed and discussed (Section \ref{sec:salientExpResults}).

\subsection{Evaluation datasets}
\label{sec:evalDs}

\newcommand{\sampleSz}[0]{30}
To construct a gold standard for this evaluation, we selected a random sample of \sampleSz{} terms from the \gls{osn} (see Section \ref{sec:osn}) and \sampleSz{} terms from the \gls{gno} (see Section \ref{sec:geonames}).
\changed{This random sample corresponds to approximately 1\% of terms in \gls{osn}, and to 4\% of terms in the \gls{gno}.}
%This sample size results in a confidence interval of about $\pm$14\% for both datasets, at a confidence level of 95\%.
The sample terms were manually mapped to semantically salient \gls{wn} synsets.
By manually selecting correct mappings between the \sampleSz{} terms from the \gls{osn} and \gls{wn} synsets, we obtained a human-generated mapping $M_h$, which includes 114 correct mappings for the \gls{osn}, and 122 mappings for the \gls{gno}.
For the purpose of replication, these test datasets are available online.\footnote{See files \url{osm_semantic_network.manual_wordnet_mapping.rdf} and \url{geonames.manual_wordnet_mapping.rdf} at \url{http://github.com/ucd-spatial/OsmSemanticNetwork}}
%These datasets can be utilised as gold standard to evaluate \gls{algo}, our semantic mapping technique.

\subsection{Evaluation measures}
\label{sec:evalMeasures} 

To evaluate the performance of \gls{algo}, we define the following performance measures (see Table \ref{table:notations} for notations).
Following \citet{euzenat:2007:semantic}, we assume that a correct mapping belongs to the machine and human mapping $m \in M \wedge m \in M_h$, \changed{while an incorrect mapping only belongs to the machine mapping, i.e. $m \in M \wedge ~m \notin M_h$.}
Hence, we define precision $P$ and recall $R$ of mapping $M$ as:

\begin{equation}
P_M = \frac{|M \cap M_h|}{|M|} ~~ R_M = \frac{|M \cap M_h|}{|M_h|} ~~~ P_M,R_M \in [0,1]
\end{equation}
 
\noindent As a general trade-off in the semantic mapping between the \gls{osn} and \gls{wn}, we favour precision over recall.
In other words, false negative mappings are preferred to false positives.
To combine the two measures into a single measure of performance that favours precision over recall, we use a $F$-measure, defined as:

\begin{equation}
F_{M\beta} = \frac{(1 + \beta^2) \cdot P_M \cdot R_M}{\beta^2 P_M+R_M} ~~~ \beta = .5,~ F \in [0,1]
\end{equation}

\noindent where $\beta = .5$ puts more emphasis on precision than recall.
All these measures fall in the interval $[0,1]$, with $1$ as the best possible result ($M \equiv M_h$), and $0$ as the worst ($M \cap M_h = \emptyset$).
This measures are used as indicators of the quality of the semantic mapping in the next sections. 

\subsection{Preliminary experiments with LIMES}
\label{sec:evalTools}

To verify the need for \gls{algo}, we tackled the problem of mapping between a vocabulary and \gls{wn} with existing semantic matching tools.
In particular, we performed the linkage between the \gls{osn} and \gls{wn} with the \emph{LInk discovery framework for MEtric Spaces} (LIMES), described in Section \ref{sec:relwork_opendataint}.\footnote{The experiments were conducted with LIMES v.0.6.}
Although the Silk framework \citep{volz:2009:silklinkdiscovery} provides similar functionality, LIMES was preferred because of its efficiency and the guarantee of full recall on all the possible mappings.

In order to align the \gls{osn} with \gls{wn}, several configurations of LIMES were defined.
LIMES computes potential mappings in two given datasets by combining string similarity measures on specific fields.
In this context, relevant fields to be compared are the key and value of the \gls{osm} concept (\semuri{osnp:keyLabel} and \semuri{osnp:valueLabel}).
In \gls{wn}, the fields are the synsets' definitions (\semuri{wns:gloss}) and the corresponding word senses' labels (\semuri{rdfs:label}).
The string similarity of these four fields can be used to compute the mappings.
The fuzzy string similarity function based on \emph{trigrams} was applied to the fields.
Pairs obtaining a similarity equal to or greater than a given threshold are included in the mapping.

\changed{
Using LIMES, we computed the entire mapping between 4,363 OSM concepts and 71,691 \gls{wn} noun synsets using two different strategies, one using only the concepts' labels, and one focused on the lexical definitions.
The mappings were then evaluated against the human-generated evaluation dataset, computing precision and recall for each case.}
%The results are summarised in displayed in Table \ref{table:limesResults}.
When matching OSM concepts and WordNet synsets only based on their labels (e.g. `amenity=university' and `university'), \changed{the mapping contains very few relevant synsets (max $P_M = .24$, with a similarity threshold $\geq .9$).}
This experiment also obtained low recall ($R_M < .1$), due to the lack of mappings with related terms from the lexical definitions.
As the system has no information about the semantic salience of specific word senses, all the word senses are included.
%Hence, the precision is very lo, essentially equivalent to a random selection of word senses.
%Raising the threshold towards $1$ does not improve the precision.

The other set of experiments was performed on the lexical definitions of the OSM concepts (\semuri{skos:definition}) and those of WordNet synsets (\semuri{wns:gloss}).
In this case, the mapping obtained even lower recall and precision, suggesting that a simple string similarity function applied on definitions does not capture their semantic salience.
These two experiments show that, while the basic functionality provided by frameworks such as LIMES is useful in several contexts, especially with very large datasets \citep{ngomo:2011:limeslargelink}, specific strategies such as \gls{algo} are needed to generate an appropriate mapping between a vocabulary and WordNet.
The next section details the evaluation of \gls{algo}.

%\begin{table}[t]
%%\centering
%\begin{tabular}{llrr}
%		\hline
%		 OSM fields & WordNet fields  & Recall & Precision \\
%		\hline
%		 \semuri{osnp:keyValue},\semuri{osnp:valueValue} & \semuri{rdfs:label} & $< .24$ & $< .3$ \\
%		\semuri{skos:definition} & \semuri{wns:gloss} & $< .1$ & $< .1$ \\
%	    \hline
%\end{tabular}
%\caption{Best performance of LIMES mappings against the evaluation dataset}
%\label{table:limesResults}
%\end{table}

\subsection{Experiment set-up}
\label{sec:salientExpSetup}

% experiment code: /Users/andreaballatore/TextSimilarity. 
% Command: run maplod

\newcommand{\taxSize}[0]{6,312}
\newcommand{\Ncases}[0]{396}

The algorithm \protect\gls{algo} takes five parameters: $V,W,ol_{min},f_{min},$ and $\Theta$ (see Section \ref{sec:mappingOsnToWn}). 
Keeping the vocabulary $V$ and \gls{wn} $W$ constant, we want to assess the impact of the other three parameters, $ol_{min}$, $f_{min}$, and $\Theta$.
Hence, we define the following parameters:

\begin{itemize}
  \item Salient taxonomy $\Theta$: either $\Theta \equiv W$ (i.e. taxonomy disabled), or a taxonomy of geographic terms (2 options);
  \item Minimum lexical overlap $ol_{min}$: $\{ 0,1,2,\ldots10 \}$ (11 options);
  \item Minimum word sense frequency $f_{min}$: $\{ 0,1,2,3,4,5,10,20,30, \ldots 100 \}$ (18 options);
\end{itemize}

\noindent These parameters result in $2 \cdot 11 \cdot 18 = \Ncases$ unique combinations of parameters.
A random disambiguation approach is added as a baseline. 
In order to disambiguate the terms from the \gls{osn} and the \gls{gno} to the corresponding word sense in \gls{wn} synsets, we select a subset of the \gls{wn} taxonomy $\Theta$ that is relevant to the geographic domain.
By manually observing the upper level of \gls{wn} (i.e. synsets with depth $\leq 3$), we selected eight synsets as roots of the salient taxonomy (see Table \ref{table:salientTaxonomy}).
All children synsets were subsequently recursively extracted, resulting in a salient taxonomy $\Theta$ of \taxSize{} noun synsets, navigating the \semuri{wns:hyponymOf} and \semuri{wns:partMeronymOf} relations.
The salient taxonomy corresponds to about $7\%$ of the entire \gls{wn} noun taxonomy.
The algorithm was executed on the \Ncases{} parameter combinations, parallelised in ten separate threads on both evaluation datasets.

\begin{table}[t]
\centering
\begin{tabular}{ll}
		\hline
		\multicolumn{2}{c}{Salient taxonomical roots in \gls{wn}} \\ \hline
	\semuri{wn:location-noun-1} & \semuri{wn:artifact-noun-1}\\
	\semuri{wn:land-noun-2} &\semuri{wn:activity-noun-1}\\
	 \semuri{wn:ecosystem-noun-1} & \semuri{wn:water\_system-noun-1}\\
 \semuri{wn:natural\_object-noun-1}~~~ & \semuri{wn:natural\_phenomenon-noun-1}\\
	    \hline
\end{tabular}
\caption{Salient synsets in the upper part of the \protect\gls{wn} taxonomy}
\label{table:salientTaxonomy}
\end{table}

\subsection{Experiment results}
\label{sec:salientExpResults}

The experiment generated \Ncases{} mappings of the \gls{osn} and \Ncases{} mappings for the \gls{gno}.
Each mapping was compared with the human-generated dataset described in Section \ref{sec:evalDs}, obtaining precision, recall, and $F$-measure.
In order to analyse the impact of each parameter on the results, we summarise the performance indicators in Table \ref{table:expResults}, showing the mean precision $\bar{P}_M$, recall $\bar{R}_M$, and $F$-measure $\bar{F}_M$.
%The LIMES results are the best ones obtained in Section \ref{sec:evalTools}.
Although \gls{algo} performs better on the \gls{osn} ($P= .92, R=.98, F=.92$) than on the \gls{gno} ($P= .86, R=.9, F=.71$), the results show highly consistent patterns across the two datasets.
As expected, precision and recall tend to be inversely proportional.
All of the three salience indicators ($\Theta$, $f_{min}$, $ol_{min}$) have a positive impact on precision, and negative on recall.

In the case of the \gls{osn}, the filter based on the salient taxonomy $\Theta$ improves the mean precision $\bar{P}_M$ from $.72$ to $.81$, with a minimal loss of recall.
On the \gls{gno}, the gain in precision is smaller but still detectable.
The filter based on $f_{min}$ increases the mean precision at the expense of the mean recall on both datasets, obtaining the best results when  $f_{min}=1$.
The minimum lexical overlap $ol_{min}$ has a similar effect on the performance, generating the best results when $ol_{min}=1$ and $2$.
These results confirm the validity of the key ideas behind \gls{algo}, described in Section \ref{sec:algo}, indicating that each of the three filters contributes to improve the overall quality of the mapping.   

Given that our objective is to maximise the $F_M$ score, biased towards precision, all the three filters need to be utilised in \gls{algo}.
In particular, the highest $F_M$ is obtained when the salient taxonomy $\Theta$ filter is on, the minimum frequency $f_{min}$ is 1, and the minimum overlap $ol_{min}$ is 1 for the \gls{osn}, and 2 for the \gls{gno}.
For the \gls{osn}, the selection of these optimal parameters ($\Theta$ on, $f_{min}=1$, $ol_{min}=1$) results in $P_{M}=.91$, $R_{M}=.98$, and therefore $F_{M}=.92$.
For the \gls{gno}, the best results consist of $P_{M}=.81$, $R_{M}=.45$, and $F_{M}=.7$.
\changed{These results confirm that \gls{algo} is able to generate a high-quality semantic mapping, vastly outperforming generic tools such as LIMES.}

\begin{table}[t]
% experiment code: /Users/andreaballatore/TextSimilarity. 
% Command: run maplod
% Analysis code in file wn_mapping_analysis.R

% GeoNames experiment data: wnGnMappingExpResults.txt
% VERY GOOD TRICK TO JUSTIFY TABLE! (package: graphics)
%\resizebox{\columnwidth}{!}{%
\begin{tabular}{lr|lll|lll}
		\hline
		            Parameter     &       Parameter        &   \multicolumn{3}{c|}{\emph{OSM Sem. Net.}} & \multicolumn{3}{c}{\emph{\gls{gno}}} \\

		name & value &  ~$\bar{P}$ &  ~$\bar{R}$ &  ~$\bar{F}$ &  ~$\bar{P}$ &  ~$\bar{R}$ &  ~$\bar{F}$\\
	    %\hline
	    \hline
	    %LIMES baseline & $-$ & .1 &  .24 & $-$\\
	    %\hline
	    %Overall & $-$ & .84 & .5 & .7\\
	    \hline
Salient taxonomy $\Theta$ & off & .79 & .5* & .67 & .77  & .40* & .61 \\
 & on & .88* & .49 & .73*						  & .79* & .36  & .62* \\
 \hline
Minimum word   & 
 \emph{(off)} 0 & .82 & .56* & .71& .77	&.44*	&.62\\
sense frequency 
 & 1 & .84 & .56*& .72*& .77	&.43	&.63*\\
$f_{min}$  & 2 & .84 & .54 & .71&.77	&.42	&.62\\
 & 3 & .84 & .53 & .71&.77	&.41	&.62\\
 &  &  & \ldots & & & \ldots &  \\
 %& 4 & .84 & .51 & .71\\
 %& 5 & .85 & .51 & .71\\
 %& 10 &.84 & .48 & .71\\
 & 20 &.85 & .45 &  .7 &.79	&.35	&.62\\
 & 30 &.85 & .44 &  .7 &.8	&.33	&.61\\
 & 100&.86* & .4 &  .69 &.81*	&.32	&.61\\
 \hline
Minimum lexical & \emph{(off)} 0 & .7 & .82* & .71 & .61	& .6*	& .59 \\
overlap $ol_{min}$ 
 & 1 & .75 & .81 &.75* & .65	& .59		&.62\\
 & 2 & .87 & .49 &.74  & .8		&.41	& .67*\\
 & 3 & .88 & .37 &.68  & .82	 &.3		& .61\\
 &  &  & \ldots & & & \ldots & \\
 %& 4 & .88 & .36 &.68  \\
 %& 5 & .89 & .35 &.68  \\
 %& 6 & .89 & .35 &.68  \\
 & 7 & .89 & .35 &.68 & .83		& .3	& .61 \\
 & 8 & .9* & .35 &.68 & .84*	& .3	& .61 \\
 \hline
Upper bounds & $-$ & .92 & .98 & .92 & .86 & .9 & .71 \\
 \hline 
\end{tabular}
%}
\caption{Summary of experiment results. Mean precision ($\bar{P}$), mean recall ($\bar{R}$), and mean F-score ($\bar{F}$). (*) Best results.}
\label{table:expResults}
\end{table}

% WHY GN != OSN?
This performance indicates that the \gls{algo} encountered considerably more difficulties with \gls{gn} terms than with the \gls{osn}.
By manually inspecting the mappings, it is possible to notice that, compared with the \gls{osn}, the \gls{gno} tends to contain specific and technically complex terms, such as \emph{talus slope}, \emph{salt pond}, \emph{interfluve}, \emph{cuesta}, and \emph{oxbow lake}, which are more challenging to map than common terms such as \emph{mountain} or \emph{road}, \changed{resulting in lower precision}.
Another reason that accounts for the lower recall is the fact that definitions in \gls{gn} are more concise, with an average of $10.9$ words per definition, while the \gls{osn} definitions have on average $38.8$ words.
While OSM definitions are indeed noisier than those in \gls{gn}, this case highlights that the algorithm suffers from a limited information problem when the lexical definitions are too concise.

A possible solution to mitigate this limitation and increase the recall could consist of extending the search for similar terms in \gls{wn} by visiting related terms.
% comment on quality of results
Although performance improvements are certainly possible, as is discussed in the next section, we consider these results satisfactory for the evaluation of our approach to semantic mapping \gls{algo}.
The precision, recall, and F-measures obtained by \gls{algo} are comparable with the performance of the state-of-the-art ontology alignment techniques recently evaluated in the context of the Ontology Alignment Evaluation Initiative.\footurl{http://oaei.ontologymatching.org/2012/results}
The full mapping between the \gls{osn} and \gls{wn}, performed with the optimal parameters, is available online as part of the network. %\footurl{http://wiki.openstreetmap.org/wiki/OSMSemanticNetwork}

%=============================================
\section{Conclusions}
\label{sec:concl}
%=============================================
\glsresetall

\Gls{lod} constitutes a promising paradigm to create a shared semantic space, in which heterogeneous geospatial datasets can inter-operate.
In the \gls{lodcloud}, \gls{wn} can be used as shared semantic ground to enable inter-operability between heterogeneous vocabularies.
In this paper, we described our contribution to the \gls{lod} vision.
First, we outlined a semantic mapping algorithm, \gls{algo}, which aims at generating semantic links between a given vocabulary and \gls{wn}.
 This algorithm offers a general semantic mapping technique between a specialised vocabulary and the well-known lexical database \gls{wn}. Given an input term from the vocabulary, \gls{algo} identifies salient synsets in \gls{wn} using three salience indicators: (1) the usage frequency of a term; (2) the term overlap between the lexical definition of the given term and the \gls{wn} definition; and (3) a manually selected salient taxonomy.
Second, we evaluated \gls{algo} on a random sample of terms from the \gls{osn}, and from the \gls{gno}, obtaining a satisfactory performance.
%We draw the following conclusions:

% Applications of OSN
\gls{algo} provides a semantic support tool to exploit \gls{lod} in geo-applications, increasing the integration of datasets at the schema level.
Using \gls{wn} as a semantic hub enables the discovery of implicit semantic relations between features, such as subsumption or meronomy, as well as the discovery of affordances, a promising approach to computational modelling the role of places.
Through federated queries over the \gls{lodcloud}, these semantic mappings can support tasks at the instance level, facilitating the matching of the same entities across \gls{lgd}, \gls{dbp}, GeoNames, and other \glspl{gkb} \citep{Ballatore:2012:survey}.

%\begin{itemize}
%  \item The \gls{osn} bridges the semantics of \gls{osm} data and the \gls{lodcloud}.
%  The network is extracted from the \gls{oww}, a repository where contributors define, edit, and document the semi-structured folksonomy of tags.
%  The dataset is structured as a SKOS vocabulary of terms utilised to describe \gls{osm} geographic features.
%  We made the \gls{osn} freely available online,\footurl{http://wiki.openstreetmap.org/wiki/OSMSemanticNetwork} and we linked it to existing semantic resources, including \gls{lgd}, \gls{dbp} and \gls{wn}.
  %The nature and quality of this mapping are discussed in Section \ref{sec:osnLod}.
Despite the advances reported in this article, our proposal for the bootstrapping of geo-vocabularies in the \gls{lodcloud} presents a number of limitations and open challenges.
\changed{\gls{wn} is a general-purpose semantic resource, and its coverage of geographic terms is limited.
While the proposed mapping technique is effective with common terms (e.g. \emph{bay}, \emph{city}, \emph{university}), it would not perform well with many technical terms in highly specialised vocabularies, such as the CORINE Land Cover of the European Environment Agency.}
As usual in the case of semantic techniques, the generated mappings contains inevitably some degree of noise, ambiguity, and incorrect semantic mappings.
SKOS mapping relations are semantically limited, and cannot express the complexity of identity relations discussed by \cite{halpin:2010:owl}.
Whether a specific semantic mapping is fit-for-purpose, depends on the application in which \gls{lod} is being used.
For example, a precision of $.8$ could be sufficient for data exploration, but could be impractical to execute complex spatial reasoning procedures.
\changed{Future work should include the comparison of other resources as semantic hubs, such as \gls{dbp} and the \gls{gno}. A larger sample of manual mappings will help evaluate the techniques more thoroughly.}
%Being a semi-structured folksonomy, the \gls{osn} does not necessarily reflect ontological commitments in the vector data, and should therefore be utilised taking into account the intrinsic uncertainty of \gls{vgi}.
%\gls{algo} was tested on the \gls{osn}, obtaining high precision ($.91$), recall ($.98$), and $F$-measure ($.92$).
  %The resulting high-quality semantic mapping is part of the \gls{osn}.
%\end{itemize}

Structuring geographic information according to the \gls{lod} paradigm provides a valuable contribution to deliver richer, more structured geospatial information to both humans and machines.
However, the \gls{lodcloud} presents a number of limitations that need to be addressed, in particular in relation to the management of identity \citep{jain:2010:linked}, and spatio-temporal reasoning \citep{janowicz:2012:geospatiallinked}.
These issues notwithstanding, the \gls{lodcloud} provides the potential for a vast, open laboratory to a growing community of scientists, software developers, and GIS specialists.
Integrating datasets with \gls{wn} is one of the avenues towards the accomplishment of that vision.

\subsection*{Acknowledgements}

The research presented in this article was funded by a Strategic Research Cluster grant (07/SRC/I1168) by Science Foundation Ireland under the National Development Plan. The authors gratefully acknowledge this support.

% END OF PAPER
%=============================================
\bibliography{../bib/thesis,../bib/mypub}
\bibliographystyle{chicago}

\end{document}